\documentstyle[prl,aps,psfig]{revtex}
\bibstyle{unsrt}

\tighten
\begin{document}
\topmargin=-1.cm
\draft

\twocolumn[\hsize\textwidth\columnwidth\hsize\csname 
@twocolumnfalse\endcsname

\title{Numerical simulation of vacuum particle production: applications to 
cosmology, dynamical Casimir effect and time-dependent non-homogeneous dielectrics}
\author{Nuno D. Antunes}
\address{Centre for Theoretical Physics, University of Sussex,
Falmer, Brighton BN1 9QH, U.K.}

\maketitle

\begin{abstract}
 We develop a general numerical method aimed at studying particle
 production from vacuum states in a variety of settings. As a first
 example we look at particle production in a simple cosmological
 model. We apply the same approach to the dynamical Casimir effect, 
 with special focus on the case of an oscillating mirror. 
 We confirm previous estimates and obtain long-time
 production rates and particle spectra for both resonant and off-resonant
 frequencies. Finally, we simulate a system with space and 
 time-dependent optical properties, analogous to a one-dimensional
 expanding dielectric bubble.
 We obtain simple expressions for the dependence of the final
 particle number on the expansion velocity and final dielectric constant.
 
\end{abstract}


\vskip2pc]
     
\section{Introduction}
\label{secI}

 One of the most intriguing and fascinating features of quantum field
 theory resides in the non-trivial nature of its vacuum states. Quantum
 fluctuations  present in the vacuum are responsible for 
 non-classical effects that can be experimentally detected.
 The most well known of such phenomena is perhaps the Casimir effect
 \cite{Casimir}. In this setting, the vacuum energy between two static
 metallic plates is shifted due to quantum fluctuations, resulting in an
 attractive force between the plates.
 In general non-static situations, changes in a system's parameters may 
 lead to a redefinition of the natural vacuum state.
 As a result, a system initially in the vacuum
 may at later stages display a  non-zero particle content.
 The possibility of this type of particle production mechanism has been
 identified and studied in a large variety of systems.
  In a cosmological context, particle creation may occur in
  back-hole radiation - the Hawking effect \cite{Hawking}, and
  as a consequence of an expanding gravitational background
  \cite{Birrell_Davies}.
 In the laboratory it is possible that dynamical generalisations of 
 the Casimir effect may lead to observable effects \cite{DodonovII}.
 Alternatively, there have  been efforts to reproduce the
 Hawking effect using specific matter systems
  \cite{Unruh,Visser}. This is the case of slow-light
  experiments \cite{Ulf}, black-hole sonic analogues in Bose-Einstein
  condensates \cite{Garay} and dielectric media \cite{Schutzhold}.

 The calculations involved in determining the physical outcome of
 particle creation processes, though trivial to state, are often
 hard or impossible to complete. Usually one is required to
 find  the solution of a set of space and/or time-dependent
 field equations, with initial conditions covering a complete
 basis of functions. Even when relying on simplifying approximations,
 the set of problems for which solutions can be found is considerably
 limited. On the other hand, linear partial differential equations
 are relatively easy to solve numerically, and the above task is within
 present computational capabilities. In this paper we explore this 
 possibility and introduce a fully numerical scheme for studying
 particle production in general settings. We will use examples
 with known solutions to help developing and testing the required
 numerical techniques. These will then be applied to new cases,
 extending original predictions and leading to new results.
 
 In Section~\ref{secII}, we rely on a simple cosmological example to 
 introduce notation and discuss concepts relevant for the rest of 
 the paper. We  use this system to describe the general numerical
 approach and we test it, by comparing the results with known
 analytical predictions.
 In Section~\ref{secIII}, we apply these techniques to the
 case of particle creation by a reflecting moving boundary in a 
 one-dimensional cavity. For an oscillating motion we
 obtain particle production rates and spectra, for both resonant
 and non-resonant frequencies, in the short and long-time regimes.
 In Section~\ref{secIV}, we study a system with time
 dependent optical properties, mimicking a bubble of dielectric material 
 expanding into the vacuum. From the numerical data we 
 determine the dependence of the final particle number on the expansion
 velocity and strength of the dielectric.
 Finally, in  Section~\ref{secV} we discuss our results and suggest further
 applications.
 Appendix~\ref{appA} contains a description of the numerical
 algorithm used in Section~\ref{secIII} to solve the field equations
 for moving boundary conditions.

 \section{A First Example: Particle Creation in an Expanding
 Background}
 \label{secII}

  In this section we will look at a simple example of particle
  creation in a cosmological system. In particular, we will study
  the evolution of a free complex field $\phi$ in an expanding
  flat universe. Here and throughout the paper we will follow the conventions and
  notation of reference\cite{Birrell_Davies}. 
     
  For a general gravitational background, the equation of motion for
  a scalar field is given by
  \begin{equation}
     (-g)^{-1/2} \partial_\mu [(-g)^{1/2} g^{\mu \nu} \partial_\nu
     \phi]+ m^2  \phi=0,
  \label{EOM}
  \end{equation}
  where $g_{\mu \nu}$ is the space-time metric and $g=|\det g_{\mu \nu}|$.

    The invariant scalar product of any two solutions of Eq.~(\ref{EOM})
  is defined as
  \begin{equation}
  (\phi_1,\phi_2)=  -i \int_{\Sigma} \left[\phi_1\,
      \partial_\mu \phi^*_2 - \partial_\mu \phi_1\, \phi^*_2\right]
      \sqrt{-g_{\Sigma}} \,d\Sigma^\mu,
 \label{scalar product}
 \end{equation}
 where the integral is taken over any space-like surface of the
 space-time ($d\Sigma^\mu$ is the time-like unit-vector orthogonal
 to the surface). Using  Eq.~(\ref{scalar product}) we can construct
 orthonormal basis of solutions of the equations of motion, obeying: 
 \begin{equation}    
   (u_i,u_j)=\delta_{ij},\,\,\,\,\,\,  
   (u_i^*,u_j^*)=-\delta_{ij},\,\,\,\,\,\,
   (u_i,u_j^*)=0 
 \end{equation} 
 Expanding the field operator in terms of a given basis we obtain the
 familiar creation and annihilation operators $a_i^{\dagger}$, $a_i$: 
 \begin{equation} 
   \phi(x)=\sum_{i}\, [a_i u_i(x) + a_i^{\dagger} u_i^*(x)]
 \end{equation}     
 Particle states are defined in respect to $a_i^{\dagger}$, $a_i$.
 In general we will be looking at systems which are static and flat
 for both early and late times - the so called {in-out} type problems.
 For these cases we can single-out in a natural way, two different
 basis of solutions, $\{u^{\rm in}_i\}$ and
 $\{u^{\rm out}_j\}$ which tend respectively as $t\rightarrow t_{\rm
 in}$  and $t\rightarrow t_{\rm out}$, to the usual flat-space positive 
 frequency modes. The field can be quantised in terms of any of 
 these two basis, leading to different definitions of creation and
 annihilation operators and hence to different definitions of particle
 states. In most cases of interest, the particle content of a given
 state will change depending on the set of modes we chose as basis.
 In particular, we will consider the case where the system is in the
 vacuum state corresponding to the {\it in} modes.
 In respect to the {\it out} particle states - the natural choice for
 late times - this state may have a non-zero particle content. The
 particle spectrum is given by \cite{Birrell_Davies}
 \begin{equation}
  N_i=\sum_j |\beta_{ij}|^2
 \label{spectrum}
 \end{equation}
 where $N_i$ is the number of particles in the mode $i$.
 The $\beta_{ij}$'s are a subset of the so-called Bogolugov
 coefficients which relate the two basis and can be
 evaluated using \cite{Birrell_Davies}:
 \begin{equation}
 \alpha_{ij}=(u_i^{\rm out},u_j^{\rm in}),\,\,\,\,\,\,
 \beta_{ij}=-(u_i^{\rm out},[u_j^{\rm in }]^*)
 \label{bog}
 \end{equation}

 In this paper, we will look into how to tackle the calculation
 of the particle  spectrum for general {\it in-out} problems,
 using a fully numerical approach. We will start by evaluating numerically
 the {\it in} modes $\{u^{\rm in}_i\}$.
 This is done by solving the equation of motion Eq.~(\ref{EOM}), taking
 as initial condition the asymptotic, flat-space mode solutions
 corresponding to the Minkowsky vacuum. These are propagated
 up to $t_{\rm out}$,  when the new stationary flat configuration
 is reached. The resulting {\it in} modes can then
 be contracted with the {\it out} modes, by evaluating numerically
 the integral Eq.~(\ref{scalar product}) at $t=t_{\rm out}$. In this
 way we obtain the Bogolugov coefficients Eq.~(\ref{bog}).
 Other physical quantities, such as the particle spectrum can be
 evaluated in terms of these.
      
 To illustrate the method, we look at a simple example of a 
 $1+1$ space-time with a time dependent scale factor.
 In conformal coordinates, the metric is given by
  \begin{equation}
   {\rm d}s^2 = a^2(\eta)({\rm d}\eta^2-{\rm d}x^2),
  \end{equation}
 and the field equations of motion are    
  \begin{equation}
   \frac{\partial^2 \phi}{\partial \eta^2} = 
  \frac{\partial^2 \phi}{\partial x^2}-m^2 a^2(\eta) \phi
  \label{EOM_expansion}
  \end{equation}
 For this system, the time independent inner product of two solutions
 is given by
 \begin{equation}
  (\phi_1,\phi_2)= -i \int_{-\infty}^{+\infty} \left[\phi_1\,
  \partial_\eta \phi^*_2 -\partial_\eta \phi_1\, 
  \phi^*_2\right]\,dx
  \label{inner_prod_conf}
 \end{equation}     
 We take $a(\eta)\rightarrow a_{\rm in}$ as $\eta\rightarrow
 -\infty$, and $a(\eta)\rightarrow a_{\rm out}$ as $\eta\rightarrow
 +\infty$.  In these limits, the corresponding mode basis have
 the usual flat space-time form:
 \begin{eqnarray}
   &&u^{\rm in}_k\rightarrow\frac{1}{2 \sqrt{\pi \omega_{\rm in}}} 
   e^{i k x - i \omega_{\rm in} \eta},\,\,\,\,\,
   \omega_{\rm in}=\sqrt{k^2+m^2 a_{\rm in}^2} 
   \label{in_modes}\\
  && u^{\rm out}_k\rightarrow\frac{1}{2 \sqrt{\pi \omega_{\rm out}}} 
   e^{i k x - i\omega_{\rm out} \eta},\,\,\,\,\,
   \omega_{\rm out}=\sqrt{k^2+m^2 a_{\rm out}^2}
   \label{out_modes}
 \end{eqnarray}
 For the scale-factor $a(\eta)$ we choose a particular form for
 which there is a known analytical solution \cite{Birrell_Davies}:
 \begin{equation}
   a^2(\eta)=\frac{1}{2}(a_{\rm out}^2+a_{\rm in}^2)+
     	\frac{1}{2}(a_{\rm out}^2-a_{\rm in}^2) \,{\rm tanh} (\rho \eta)
 \end{equation}
 Taking Eq.~(\ref{in_modes}) as initial condition, we evolve the
 equation of motion Eq.~(\ref{EOM_expansion}) up to a time when
 $a\sim a_{\rm out}$. The numerical solutions are then contracted with
 the {\it out} modes Eq.~(\ref{out_modes}),
 using Eq.~(\ref{inner_prod_conf})  
 to obtain the Bogolugov coefficients. For our choice of
 scale-factor these can be evaluated analytically. The derivation 
 can be found in \cite{Birrell_Davies}, the result being given by
  \begin{eqnarray}
  |\alpha_{k k'}|^2= \delta_{k k'}
  \frac{\sinh^2[\pi (\omega_{\rm out}+\omega_{\rm in})/(2\rho)]}
 {\sinh(\pi \omega_{\rm in}/\rho) \sinh(\pi \omega_{\rm out}/\rho)}\\
 |\beta_{k k'}|^2=  \delta_{k k'}
  \frac{\sinh^2[\pi (\omega_{\rm out}-\omega_{\rm in})/(2\rho)]}
 {\sinh(\pi \omega_{\rm in}/\rho) \sinh(\pi \omega_{\rm out}/\rho)}
 \end{eqnarray}   
     
\begin{figure}
 \centerline{\psfig{file=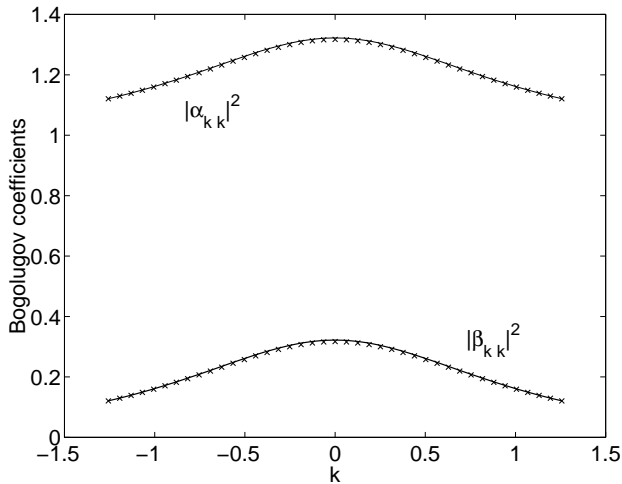,width=3.25in}} 
 \caption{Analytical (line) and numerical results (crosses) for 
  Bogolugov coefficients, for a model with  
  $m=1$, $\rho=10.$,  $a_{\rm in}=1.$ and $a_{\rm
  out}=10$.}  
 \label{expand_univ}
 \end{figure}

  In Fig.~\ref{expand_univ} we compare the numerical Bogolugov coefficients
 with the expected analytical values, for a particular choice of
 parameters  $m=1$, $\rho=10.$, $a_{\rm in}=1.$ and $a_{\rm out}=10.$  
 We evolved $40$ modes for a period of time $t_{\rm out}-t_{\rm
 in}=0.6$,  enough for the scale factor to change
 from its initial to final value with good accuracy. The equation
 of motion Eq.~(\ref{EOM_expansion}) was discretized using a standard 
 Leap-Frog algorithm, in a box of physical size $L=100$ with periodic
 boundary conditions. The lattice spacing and time-step were set to
 $dx=0.25$ and $dt=0.0001$.
 The numerical data matches very well the analytical prediction,
 confirming the accuracy of the numerical approach for the chosen
 set of parameters.
   
 Note that only diagonal elements of the Bogolugov matrix are
 non-zero, a consequence of the invariance of the system under 
 spatial translations. This implies that the particle spectrum
 coincides with the diagonal of the Bogolugov matrix, {\it i.e.}
 $N_k=|\beta_{k k}|^2$. Since Eq.~(\ref{EOM_expansion}) is
 separable in space and time, we would have the choice of evolving
 independently each mode's amplitude in Fourier space. 
 Numerically this approach would have been considerably simpler. Our
 method is 
 crucial though, in situations where this option is not available.     
 In the next sections we will apply it to a few of such cases,
 systems that have both explicit space and time-dependency.

\section{Particle Production by a Moving Mirror in a Cavity}
\label{secIII}

 In this section we will look at particle production by a moving
 mirror in a closed cavity. This system has attracted the attention
 of several authors, displaying many complex and non-trivial
 features despite its apparent simplicity. More importantly, it is the
 most natural setting for an experimental observation of 
 particle creation from a vacuum state (see \cite{DodonovII} and
 references therein). 
     
 We will consider a scalar field confined to a one-dimensional
 box with one of the boundaries fixed at $x=0$ and the other moving 
 according to a given trajectory $X(t)$ (with $|\dot{X}(t)|<1$).     
 For the
 sake of simplicity, and in order to take advantage of conformal
 invariance we assume the field to be massless, its equation of motion 
 being given by
  \begin{equation}
   \frac{\partial^2 \phi}{\partial t^2} =
   \frac{\partial^2 \phi}{\partial x^2}
  \label{EOM_moving_mirror}
  \end{equation}    
 The boundary conditions for the field are obtained by constraining it to
 vanish on both walls at all times, {\it i.e.}
 $\phi(0,t)=\phi(X(t),t)=0$. By doing so we assume the walls
 to be perfect reflectors. 
     
  In the case of a stationary mirror, time translational invariance
 makes it possible to find a set of mode solutions with positive
 definite frequency. These are given by
 \begin{eqnarray}
 && u_n(x,t) = \frac{1}{\sqrt {n \pi}} e^{-i \omega_n t} 
 \sin(\omega_n x),\nonumber \\
 &&\,\,\,\,\,\omega_n=\frac{n \pi}{L}\,\,\,\,\,\,\,\,n=1,\dots,
 \infty
 \label{basis_static}
 \end{eqnarray} 
  where $L$ is the length of the static box. These modes and their 
 complex conjugates form a complete set of solutions
 of Eq.~(\ref{EOM_moving_mirror}), orthonormal in respect to the
 simplectic product
 \begin{equation}   
  (\phi_1,\phi_2)= -i \int_0^{L} \left[\phi_1\,
      \partial_t \phi^*_2 - \partial_t \phi_1\, \phi^*_2\right]
      \,dx
 \label{scalar product cav}
 \end{equation} 
     
 The stationary solution allows us to define a general class
 of {\it in-out} problems. We will consider mirror
 trajectories such that $X(t)$ is a constant $X_0$ for $t<0$, evolves
 arbitrarily for a period of time, going back to rest at $X_f$ for $t>t_f$.
 The time and space dependence introduced in the system by the
 motion of the mirror, leads to particle production without the
 need to consider additional external fields. This phenomenon
 is known as the dynamical Casimir effect.
 
 In order to determine the final particle content of the box for a
 given $X(t)$, one must find a set of solutions for the equation
 of motion, using Eq.~(\ref{basis_static}) as initial condition. This
 problem can be formally solved by taking advantage of the conformal
 invariance 
 of Eq.~(\ref{EOM_moving_mirror}).  G. Moore showed \cite{Moore}
 (see also \cite{Fulling_Davies}) that by changing  coordinates, the 
 general problem can be mapped into the stationary case. In this way 
 a complete set of solutions can be found, with the general form 
 $u_n(x,t) \propto \exp[-i n \pi R(t-x)] - \exp[-i n \pi
 R(t+x)]$. $R(u)$ is a
 function to be determined in terms of the particular motion of
 the mirror, according to \cite{Moore}
 \begin{equation}
  R[t+X(t)]=R[t-X(t)]+2
 \label{moore_eq}
 \end{equation}
 This equation is of course non-trivial to solve in generality. One
 approach is to find $R(u)$ in terms of $X(t)$ via a perturbative 
 expansion in the velocity of the mirror \cite{Moore}.
 This technique is valid only in the adiabatic limit $\dot{X}<<1$ 
 and gives no meaningful results in the relativistic
 regime. Only for a few specific cases of the mirrors's
 trajectory have analytical solutions been found
 \cite{Cast_Ferr,Sarkar,Dodonov,Law1}. 
   
  Using the numerical procedure outlined in Section \ref{secII} the
 problem can be solved for any given arbitrary mirror trajectory. 
 The known analytical solutions will be  useful in providing tests for 
 the algorithms and in gauging the numerical requirements for
 good accuracy. 
 The general procedure will be as before: the {\it in}-modes 
 Eq.~(\ref{basis_static})
 are evolved numerically up to the final time $t_{\rm  f}$, when the
 result is contracted with the basis of stationary modes for the final
 length $X_{\rm f}$.
 The main difficulty lies in obtaining a numerical solution for the
 wave equation with moving boundaries. Since the size of the box
 changes with time we are forced to vary the number of points in the
 simulation lattice. These discontinuous jumps in the lattice size
 introduce errors in the numerical solution. We overcame
 the problem by developing an improved Leap-Frog algorithm that
 smoothes the values of the field derivatives on the boundary - a detailed
 discussion of the method can be found in Appendix \ref{appA}.
     
 Finally we should stress that although the focus here will be on one 
 dimensional massless field theories, the
 same methods can easily be extended to higher dimensions and massive
 fields \cite{preparation}. In those cases very few results are known
 \cite{mazzitelli} since one cannot rely on conformal invariance to
 simplify the
 problem. For those systems, a numerical approach may be the only
 way to tackle accurately the long time  regime.

 \subsection{Uniformly Moving Mirror}
   
 Here we will look briefly at a simple type of trajectory for which the
 {\it in-out} problem can be solved analytically
 \cite{Moore,Fulling_Davies,Cast_Ferr}, namely the case when the
 mirror moves with constant velocity. The mirror evolves from rest at
 $t=0.$  moving according to $X(t)=X_0+v t$ up to final time $t_f$, when
 it becomes stationary again. The function $R(u)$ obeying Eq.~(\ref{moore_eq})
 for this trajectory was fully determined by Castagnino and Ferraro 
 \cite{Cast_Ferr}. These authors showed that $R$ is a piece-wise
 linear function with derivative discontinuities. This is a consequence of the 
 discontinuous change in the velocity of the mirror for initial and
 final times. The final spectrum for created particles was
 estimated to be of the form  $N(k) \propto v^2/k$, in the limit of large
 $k$ and velocities much lower than the speed of light.
     
\begin{figure}
 \centerline{\psfig{file=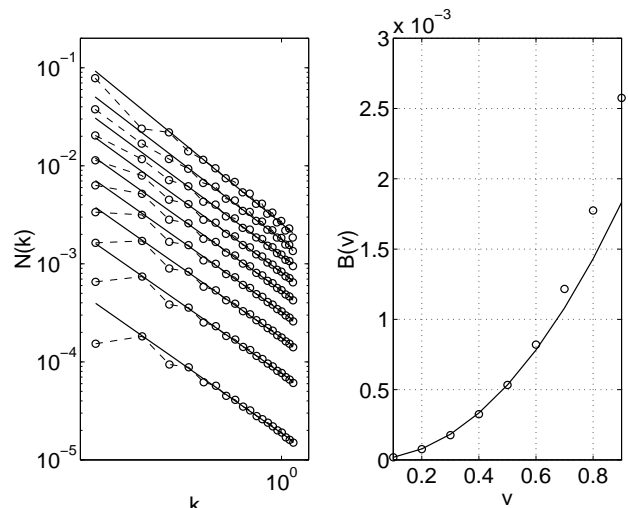,width=3.25in}} 
 \caption{Left plot: particle spectrum for mirror velocities
 varying from $0.1$ to $0.9$ (bottom to top). The straight lines
 are fits to a power law $N=B k^\alpha$, using the 15 higher momentum data 
 points.
 In the right plot we have the dependence of $B$ on the velocity.
 For low-$v$ this is well described by $B=v^{2.1}$, the
 numerical power being obtained by fitting the first $6$ data points.}   
 \label{part_num_01}
 \end{figure}

 In Fig.~\ref{part_num_01} we confirm this estimate. The results
 were obtained by simulating a box with initial physical length 
  $X_0=50.$ expanding with constant speed $v$ for a total time
 $t_f=50.$ We evaluated
 the final particle spectrum for $20$ modes, for values of $v$ ranging from
 $0.1$ to $0.9$. As seen in the figure, the spectrum for high $k$ is
 well fitted by a power law of the form $N(k)=B k^\alpha$, with $B$ and $\alpha$
 depending on $v$.  As expected, $\alpha$ is close to $-1$, varying
 between $-1.1$ for $v=0.1$ and $-1.3$ for $v=0.9$. $B$ also
 depends quadratically on the expansion velocity for small $v$. When
 relativistic speeds are reached (about $v\sim .7$) this simple
 relation stops holding, with $B$ increasing faster and leading to a
 stronger rate of particle production. 
     
 Note that the $1/k$ spectrum leads to a logarithmically divergent
 total particle number. This divergence is 
 a consequence of the discontinuous changes in the velocity of the
 mirror \cite{Moore}.
 As a rule
 of thumb we may expect particle production to be related to the
 mirror's acceleration. As this diverges for the initial and final
 times, so does the total particle number. In realistic cases however,
 the divergence should disappear, as the 
 acceleration remains bounded for all times.

 Finally let us note that for a box that expands
 continuously with constant $v$, with no initial or final
stationary points,  
 an exact solution for Eq.~(\ref{moore_eq}) can easily obtained \cite{Moore}:
 \begin{equation}
 R(u)=\frac{1}{\arctan{v}}\log\left(u + \frac{X_0}{v}\right)
 \end{equation}
 with  $X_0$ the position of the mirror at $t=0$.
 The corresponding normalised mode basis is given by     
 \begin{eqnarray}
 && v_n = \frac{1}{2 \sqrt {n \pi}}\left[
     e^{-i \Omega_n \log(X_0/v+t-x)} -
     e^{-i \Omega_n \log(X_0/v+t+x)}\right],\nonumber
 \\
 && \,\,\,\,\,\Omega_n=\frac{n \pi}{{\rm arctanh}(v)}\,\,\,\,\,\,\,\,n=1,\dots,
 \infty
 \label{basis uniform}
 \end{eqnarray} 
 This set of exact analytical solutions proved useful in testing 
 the accuracy of
 the numerical algorithm for the moving boundary problem, as detailed in 
 Appendix \ref{appA}. An interesting feature of the solution is that
 for $t_r= 2 X_0/(1-v)$
 the evolved modes coincide with the original ones, re-scaled to the
 new box length. This feature is also present for the
 {\it in-out} motion
 studied above. It implies that  at $t_r$ the {\it in} modes are
 identical to the {\it out} modes and the Bogolugov coefficients should vanish. 
 As a consequence we should observe no particle production for this 
 particular stopping time. We checked that this `resonance' does
 indeed take place by performing the simulation described above with
 final time set to $t_f=t_r$. For several choices of expansion
 velocity (since $t_r$ grows with the inverse of $1-v$ leading to very
 long simulation times, we restricted ourselves to the low-$v$ regime)
 the total number of particles produced was indeed zero.

\subsection{Oscillating Mirror}

 Moving a step up in complexity from the uniformly moving mirror
 problem, we will consider in this section a cavity whose boundary 
 executes small, periodic oscillations. This type of 
 dynamical Casimir effect setting has been widely studied, and
 there is hope that an equivalent of a vibrating mirror could be
 experimentally realized. In the particular situation where the mirror
 frequency resonates with one of the cavity modes, particle
 production effects may be magnified, leading eventually to
 observable results.
 A thorough review of the topic with extensive references can be
 found in \cite{DodonovII}.

 We will consider a sinusoidal mirror trajectory given by
 \begin{equation}
  X(t)=X_0+\frac{A}{2} \left[ 1-\cos(\omega t) \right]
 \end{equation}
 In the simulations described below, the amplitude of
 the motion $A$ will be taken to be considerably smaller than $X_0$, 
 the initial size
 of the box. Typically we will set $X_0=50$ and $A=2$.
 The frequency of the motion $\omega$, will vary for
 different runs, care being taken that the velocity of the mirror never
 exceeds the speed of light. In general $\omega$ will be set to
 be one of the natural frequencies of the cavity, $\omega_n= n
 \pi/X_0$.
 This choice of frequencies is usually favoured in the literature,
 since it is expected that resonance between the cavity modes and
 the motion of the mirror will lead to higher particle production for
 long times. 
 The resonant case is also considerably easier to tackle analytically 
 (see \cite{DodonovII} where the problem is solved exactly for $n=2$
 and \cite{cole} for a calculation of $R(u)$ for a family of 
 periodic-like trajectories). In fact, as far
 as we are aware, off-resonant solutions are only discussed in
 \cite{Pawel}, where the energy
 production is evaluated for the general case.
 Our method of course, does not distinguish between resonant and
 off-resonant frequencies - in general we will present results for
 resonant frequencies and discuss how they change for the general case. 
     
\begin{figure}
 \centerline{\psfig{file=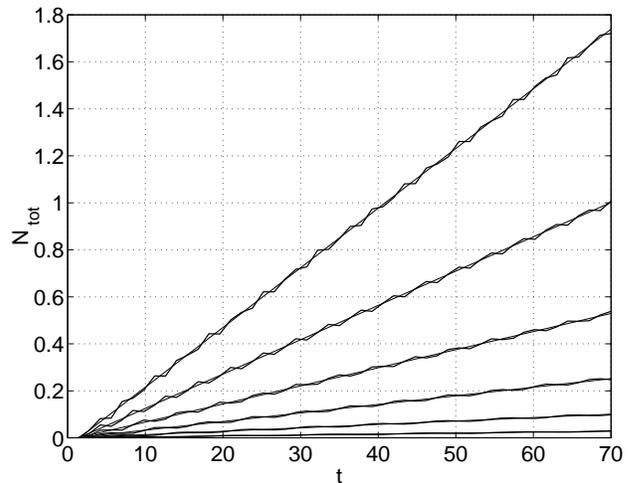,width=3.25in}} 
 \caption{Total number of particles versus time for different
 frequencies of the oscillating mirror. The straight lines are linear
 fits to $N_{\rm tot}(t)$. The oscillation frequencies are of the
 form $\omega=n\pi/X_0$ with $n=4,6,8,10,12,14$. 
 $60$ modes were used to calculate the total particle number.} 
 \label{num_vs_t}
 \end{figure}

 We start by looking at results for very short
 times, when the mirror executes a small number of oscillations.   
 In Fig.~\ref{num_vs_t} we show the total number of particles produced
 as a function of time, for several choices of the oscillation
 frequency. At every time-step, the numerical field profiles were
 contracted with the static basis for a stationary box of length $X(t)$.
 As discussed in the previous section this will lead to a convergent
 particle spectrum, but only for times when the mirror velocity is zero.
 For the sake of clarity, and with this caveat in mind, we still plot
 the total particle number for all time steps. For the number of modes
 simulated, the results for non-stationary final states do not deviate
 much from the finite case. Since the divergence is logarithmical,     
 we presume it would only manifest itself if much higher frequencies
 were included.
     
 It is clear from Fig.~\ref{num_vs_t} that the evolution is
 effectively linear for the period shown. 
 The production rate changes
 only for times of the order of twice the box length -
 up to then each oscillation produces roughly the same increase in particle
 number. This suggests that for small times, the field disturbances
 radiate away from the moving boundary without significant interference.
 For later times, after reflecting back on the stationary wall, they 
 will interact with the vibrating mirror, constructively or otherwise
 depending on their frequency, leading to a change in the 
 production rate.
     
\begin{figure}
 \centerline{\psfig{file=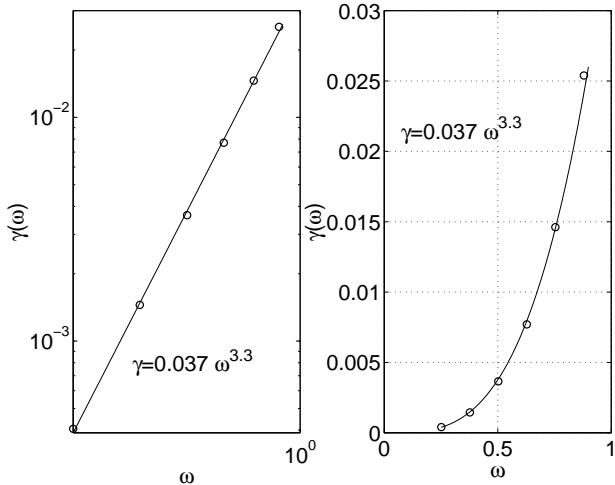,width=3.25in}} 
 \caption{Fit of the particle production rate to a power law,
 log-log and linear scales.} 
 \label{power}
 \end{figure}

 For each value of the oscillating frequency $\omega$, we fitted the 
 results to a linear law of the form
 $N_{\rm tot}(t)=\gamma t$. We checked that allowing for a 
 constant term in the linear expression, $N_{\rm tot}(t)=b+\gamma t$ 
 did not change the estimate for the production rate $\gamma$ 
 ($b$ was always found to be
 very small, of order less than $10^{-3}$).    
   Clearly, higher oscillating frequencies lead to higher particle
 production rates. In Fig.~\ref{power} the dependence of $\gamma$ 
 on the mirror's frequency is made more explicit. 
 The results are very well fitted by a power law of the form 
 $N_{\rm tot}(t) = D\, \omega^k\, t$ with $k=3.3$,
 a result in agreement with the analytical estimates in the
 literature. It is generally predicted that for a
 single oscillation the number of particles produced varies as a
 power law of the oscillation frequency. The particular value of the
 power depends on the specific type of oscillating trajectory. In 
 \cite{Cast_Ferr} 
 the production power for half an oscillation was shown to vary
 between $2.$ and $4.$, and in \cite{Sarkar} a $\omega^4$ dependence 
 was found in the limit of small velocities. For a single oscillation,
 our result corresponds to a power of  $2.3$, which is in line with these 
 values.
                  
 \begin{figure}
 \centerline{\psfig{file=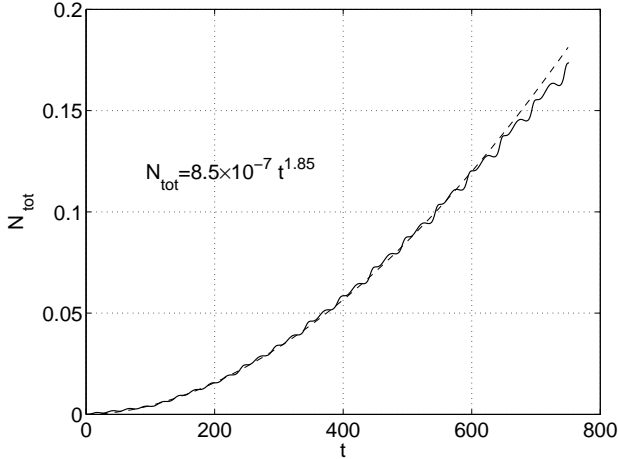,width=3.25in}} 
 \caption{Total particle production versus time for the n=2 
 resonance frequency. The fit to a power law - dashed line - is for times
  varying from $t=100$ till $t=700$. 40 modes were used in the simulation.} 
 \label{part_1}
 \end{figure}

 As for the amplitude coefficient $D$, it was suggested in
 \cite{Sarkar} that it should vary quadratically with the amplitude
 of the vibration of the mirror. We performed a quick test of
 this estimate by repeating the same set of runs with $A$, the
 mirror oscillating amplitude set to $1.$, half its previous value.
 Fitting both sets of data simultaneously to the same power $k$
 and allowing $D$ to vary, we obtained $k=3.2$, $D=0.0342$ and
 $D=0.00780$ for $A=2.$ and $A=1.$ respectively. The rate of
 the two amplitudes is $0.23$, consistent with a reduction by a factor
 of $4$ as expected.

  We now look at the longer time behaviour of the system. As 
 mentioned above, the evolution stops being
 described by a linear law  at about $t\simeq 2 X_0$, when a new
 regime with a different production rate sets in. In Fig.~\ref{part_1}
 we can see the total particle number for longer times for the
 resonant frequency $\omega=2\pi/X_0$.
     
 The evolution for the period of time shown is well fitted by
 a power law $N_{\rm tot}=8.5\times 10^{-7} t^\alpha$, with
 $\alpha=1.85$. For the dynamical range in question a power law
 growth with $\alpha=2.$ was predicted in ref. \cite{DodonovII}. This
 is in reasonable agreement with our numerical estimate. 
 
\begin{figure}
 \centerline{\psfig{file=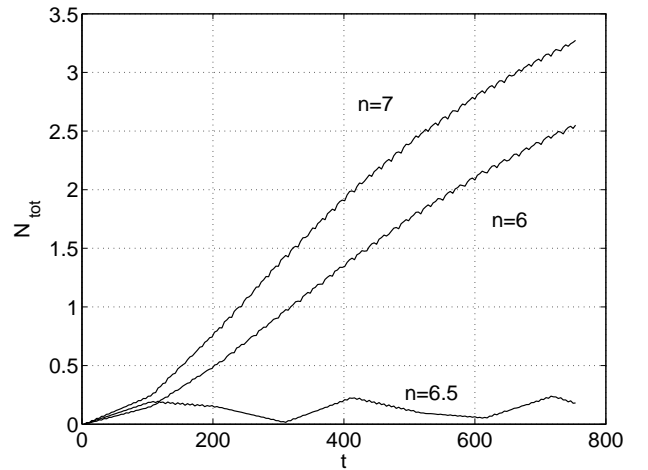,width=3.25in}} 
 \caption{Total number of particles produced for resonant frequencies
 n=6 and n=7
  and off-resonant frequency n=6.5. 40 modes used in the simulation.} 
 \label{part_3}
 \end{figure}

  For higher frequencies, as far as we are aware, there are no
  explicit  predictions for the  particle number time dependency
  (see \cite{cole,Pawel} for a calculations of the energy produced). In 
  Fig.~\ref{part_3} we show $N_{\rm tot}$ as a function of time for
  two higher resonant frequencies $\omega=6\pi/X_0$
  and $\omega=7\pi/X_0$. We also include
  $N_{\rm tot}$ for the off-resonant case $\omega=6.5\,\pi/X_0$.
     
 The overall evolution does not fit any simple expression in any of
 the three cases.
 We note that the initial period of linear evolution is always present 
 and we have checked that the production rates follow the power
 law shown in Fig.~\ref{power}, even in the off-resonant case.
 For later times though, the contrast between off-resonant and
 resonant trajectories is very clear, with limited quasi-periodic
 production in the former case and continuing increase in the later.
 It is interesting not to find any qualitative difference between the
 long time behaviour of the even and
 odd resonant frequencies systems, n=6 and n=7. In fact it has been often
 suggested that the main phenomena governing the evolution in the long 
 time limit is akin to parametric resonance \cite{DodonovII}. If that
 were strictly the case though, resonant amplification would only take place
 for oscillating frequencies equal to twice the cavity's natural
 frequencies \cite{Landau}. That is, using our notation, we should
 see long time suppression of the production rate for odd values
 of $n$, which clearly is not the case.
 This seems to confirm the results in \cite{Pawel} where it was
 found that the energy produced grows exponentially for long times,
 for any motion with $\omega$ equal to a multiple of the fundamental
 frequency $\pi/X_0$.
     
 \begin{figure}
 \centerline{\psfig{file=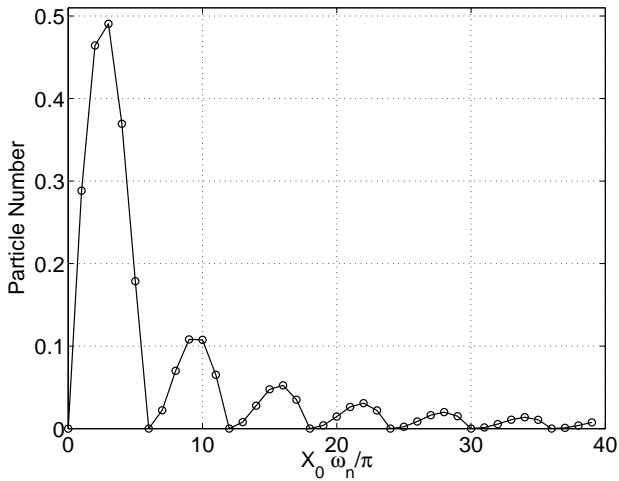,width=3.25in}} 
 \caption{Particle spectrum for $\omega=6\pi/X_0$, $t=720$.} 
 \label{spectrum1}
 \end{figure} 
     
 We now look at the distribution of the particles produced
 in terms of their frequency.
 In Fig.~\ref{spectrum1} we show the late time particle spectrum
 for the resonant $n=6$ motion.  
 We observe an alternating succession of excited and suppressed 
 frequencies. In particular we see that there is no production
 at all for frequencies multiple of the oscillating frequency.
 This cancellation is exact and it coincides strictly with points
 of the form $N \omega$, with $N$ integer. The peaks of the spectrum 
 follow a similar but less precise pattern. The most excited frequency
 is $\omega/2$ and the following peaks take place at about points
 $M \omega+\omega/2$, though there seems to be a tendency for the peak to
 move up for higher frequencies.
 The fact that the maximum of the production rate corresponds
 to a frequency of 
 about half the excitation, with higher even multiples suppressed
 and odd ones enhanced, is once again reminiscent of parametric
 resonance. Nevertheless, since the same pattern is observed for oscillating
 frequencies odd multiples of $\pi/X_0$, we must conclude that a
 more complex process is operating in the system.

\begin{figure}
 \centerline{\psfig{file=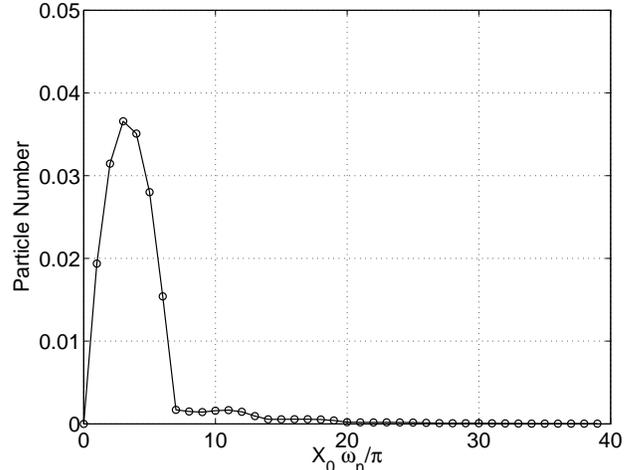,width=3.25in}} 
 \caption{Particle spectrum for the off-resonant motion $\omega=6.5\pi/X_0$,
 for $t=720$.} 
 \label{spectrum2}
 \end{figure}

  Finally we contrast the resonant spectrum with 
 the $n=6.5$ off-resonant case as shown in Fig.~\ref{spectrum2}.
 Clearly, 
 the overall amplitude is greatly reduced (by about one order of
 magnitude) and the
 pattern of alternating peaks is no longer present. The single peak
 is once again centred around $\omega/2$ but its height oscillates
 widely in time, leading to the periodic fluctuations in total particle
 number shown in  Fig.~\ref{part_3}.
     
\section{Non-uniform time-varying Dielectric}
\label{secIV}
     
 In the examples discussed above, the varying geometry 
 of the systems considered was the primary cause behind particle
 production from the initial vacuum state. In this Section we will
 look at a system with fixed dimensions, but whose bulk physical properties 
 change both in space and in time. In particular we will consider wave 
 propagation in a background with a space and time dependent dielectric 
 ``constant''. The system will be modelled  by a straightforward
 generalisation of the wave equation \cite{Law}
   \begin{equation}
  \frac{\partial}{\partial t}
  \left(\epsilon(x,t) \frac{\partial \phi}{\partial t}\right) = 
  \frac{\partial^2 \phi}{\partial x^2 },
  \label{EOM_die}
 \end{equation}	  
 where $\epsilon(x,t)$ is the varying dielectric term. Other 
 generalisations are possible, their form depending
 ultimately on the properties of the particular system one chooses to model.
 In \cite{Claudia}, the change in $\epsilon$ is caused by a
 moving half-infinite dielectric material, and compliance with
 Lorentz invariance leads to a wave equation with a cross derivative
 term. In so-called slow
 light experiments \cite{Ulf}, based on electromagnetically-induced
 transparency, the optical properties of the medium are controlled by an 
 external source. These systems can be described
 by an effective Lagrangian that includes a mass-like term for
 the field, leading to yet another choice of field theory.
 Here, for the sake of simplicity we follow \cite{Law} and 
 use Eq.~(\ref{EOM_die}), though our approach can be easily
 extended to the cases mentioned above.
     
  For two solutions of Eq.~(\ref{EOM_die}) the time invariant scalar
 product is given by
 \begin{equation}   
 (\phi_1,\phi_2)=
  -i \int_{0}^{L} \epsilon(x,t) \left[\phi_1\,
      \partial_t \phi^*_2 - \partial_t \phi_1\, \phi^*_2\right]
      \,dx
 \label{scalar product cav}
 \end{equation}
 the fields  being defined in a finite box of size $L$. We will 
 look at cases where $\epsilon(x,t)$ is spatially
 uniform both for the initial and final times, $t_1$ and $t_2$. The
 corresponding normalised mode basis are given by:    
 \begin{eqnarray}
 && u_n(x,t) = \sqrt{\frac{c_i}{n \pi}} e^{-i \omega_n t} 
 \sin(k_n x),\nonumber \\
 &&\,\,\,\omega_n=c_i k_n,\,\,\,\,k_n=\frac{n \pi}{L}\,\,\,\,\,\,\,\,n=1,\dots,
 \infty
 \label{basis static}
 \end{eqnarray}   
 where $c_i=1/\sqrt{\epsilon_i}$ with $i=1,2$ denotes the
 initial and final speed of light respectively.
    
  In the specific scenario we will be simulating, the original
 medium with dielectric constant $\epsilon_1$ is replaced throughout
 the box by a medium characterised by $\epsilon_2$. Medium $1$
 expands into medium $2$ from left to right with a given velocity
 $v$, as a one-dimensional analogue of an expanding bubble. 
 At intermediate time steps, the two regions with different dielectric 
 constants are separated by a ``wall'' that interpolates between 
 $\epsilon_1$ and $\epsilon_2$. We model the wall profile
 as a sine
\begin{equation}
\chi(x)=\left\{
         \begin{array}{cl}
         \epsilon_{1}, & \mbox{if  $x>d/2$}  \\
         \epsilon_{+} - \epsilon_{-} \sin(\pi x/d), & \mbox{if $-d/2<x<d/2$} \\
         \epsilon_{2}, & \mbox{if  $x<-d/2$} 
         \end{array}
        \right.  
\end{equation}
 with $\epsilon_{+}=(\epsilon_{1}+\epsilon_{2})/2$,
 $\epsilon_{-}=(\epsilon_{1}-\epsilon_{2})/2$, $d$ being the wall
 thickness. Using the wall profile $\chi(x)$ we can easily define
 a time-dependent dielectric function modelling the setting described above,
 as $\epsilon(x,t)=\chi(x-v t)$ with
 $t$ varying between $t_1=-d/(2\,v)$ and $t_2=d/(2\,v)+L/v$. We will
 be looking at cases where the final speed of light is less than
 the original one, corresponding to a vacuum box being filled with
 a denser medium. The initial dielectric constant will be 
 set to $\epsilon_1=1$ with $\epsilon_2>\epsilon_1$.
 In general we will
 constrain the wall expansion velocity to be less than the speed of
 light in both the initial and final medium, that is
 $v<1/\sqrt{\epsilon_2}<1/\sqrt{\epsilon_1}$  
      
\begin{figure}
 \centerline{\psfig{file=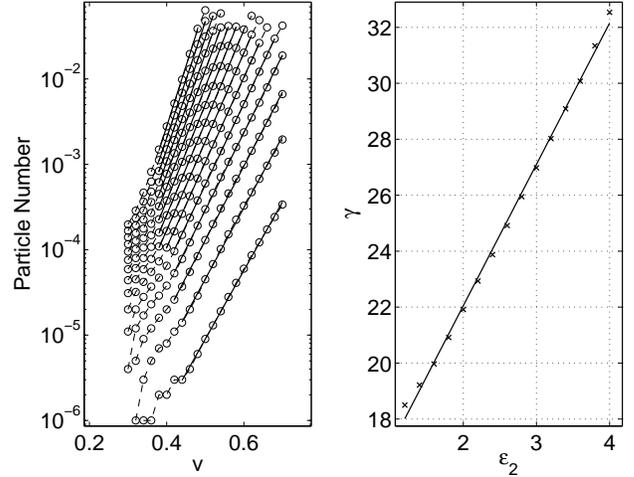,width=3.25in}} 
 \caption{Left plot: particle number {\it vs} expansion velocity for
 values of $\epsilon_2$ varying between $1.2$ and $4.0$ (bottom to
 top curves). The data was fitted for medium and high-$v$ to an
 exponential $A e^{\gamma v}$. The dependence of $\gamma$ on the
 final dielectric constant $\epsilon_2$ is shown in the right plot. The results
 are well fitted by a linear function $\gamma=\frac{K_0}{c_1}+\frac{K_1}{c_1}
 \frac{\epsilon_2}{\epsilon_1}$, with $K_0=12.0$ and $K_1=5.1$ } 
 \label{scaling}
 \end{figure}

 In Fig.~\ref{scaling} we illustrate the dependence of the number
 of particles produced on the expansion velocity. The wall thickness
 was set to $d=5.$, considerably smaller than the
 physical box size, $L=50.$ The initial dielectric
 constant was fixed to $\epsilon_1=1.$ whereas its final value 
 $\epsilon_2$  was varied between $1.2$ and $4.0$.  
 For each particular choice of $\epsilon_2$ we simulated a series
 of expansion velocities ranging from $v\simeq 0.3$ up to the 
 limit $v\simeq 1/\sqrt{\epsilon_2}$.

     We found that for low velocities the production
 rate was considerably suppressed. As the expansion velocity increases, 
 the total number of particles produced goes up steeply - on average
 by more than two orders of magnitude - as shown in  Fig.~\ref{scaling}. 
 For mid and high values of the velocity, the dependence
 of the final particle number on $v$ is well described by an exponential
 law,  $N_{\rm tot}= A e^{\gamma v}$.
 This expression is valid up to values of $v$ considerably near the speed of 
 light in the second medium.
     
 \begin{figure}
 \centerline{\psfig{file=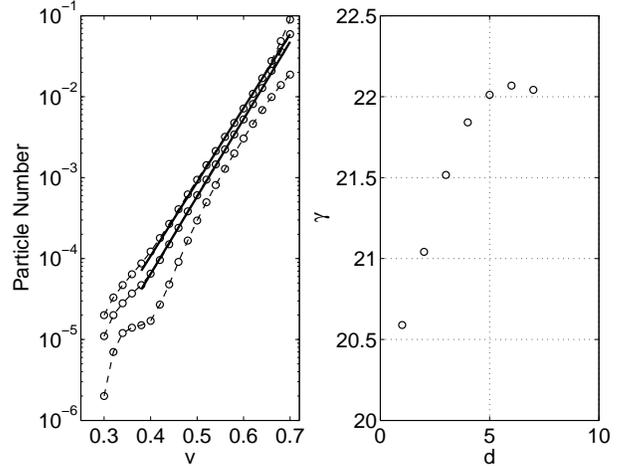,width=3.25in}} 
 \caption{Left plot: particle number dependence on the expansion velocity,
  for three different values of the wall thickness $d=2,5,12$
  (top to bottom). For the two lower values of $d$ we fitted the data to
  $N_{\rm tot}= A e^{\gamma v}$. For the higher value of the thickness,
  the results no longer follow a simple exponential law. In the right
  plot we show $\gamma$ as a function of the wall thickness for
  $d$ varying between $1$ and $7$. } 
 \label{diel}
 \end{figure} 
 
 When the final dielectric constant is allowed to vary we observe
 as expected, that stronger changes in the optical properties
 of the system lead to higher production rates. As a consequence
 we should observe an increase in the coefficient $\gamma$  for higher
 values of $\epsilon_2$. This is fact the case, and it turns out that
 the dependence of $\gamma$ on the final dielectric constant is very simple.
 In the regime studied $\gamma$ varies linearly with $\epsilon_2$ as
 can be seen in the fit shown in the right plot of Fig.~\ref{scaling}.
 Using dimensionality arguments we can then  write
 \begin{equation}
 N_{\rm tot} \propto \exp\left[
  \left(K_0+K_1 \frac{\epsilon_2}{\epsilon_1}\right) 
  \frac{v}{c_1}\right]
 \end{equation}
 where the dimensionless constants in the exponential are given by
 $K_0=12.0$ and $K_1=5.1$ for the system in question.  
     
 We also tested how the properties of the system depend on the
 thickness of the wall separating the two media. 
 For a fixed value
 of the final dielectric constant $\epsilon_2=2.$ we allowed $d$
 to vary. A sample of the results is shown in Fig.~\ref{diel}.
  As expected, for larger values of $d$ the transition between the
 initial and final media is smoother, and the production rate is
 lower. As before, 
 for $d\leq 8$ the final particle number depends exponentially on the
 expansion velocity, though for higher values of the wall thickness
 we observe deviations from this simple law. In the ``thin wall'' regime,
 the exponent $\gamma$ varies very little with $d$, as can be seen in
 the right plot in Fig.~\ref{diel}. For $1\leq d \leq7$ the change in 
 $\gamma$ is less than 10\%.
     
 Finally we checked whether these results depend on the specific
 shape of the wall profile. We replaced the sin wall by one where
 a linear function interpolates between the two dielectric constants.
 We observed that the results were virtually unchanged.
          
 \section{Conclusions}
 \label{secV}

 We have developed and tested the numerical techniques necessary
 to tackle the problem of particle production from a vacuum state in 
 general scenarios. In particular, we have shown that the approach
 performs quite well when applied to cosmology, to the dynamical 
 Casimir effect and to systems with complex optical properties. 
 In some of these cases we were able to use the numerical data to obtain
 simple phenomenological laws. This illustrates the power of the method and
 shows how it can be used as a tool to approach problems for which there
 is little or no analytical information. The opportunities are
 many, as was mentioned in the bulk of the paper. In the context of the
 dynamical Casimir effect, the first obvious step will be to generalise
 this approach to higher dimensions, where little is known analytically.
 This poses no obvious numerical difficulty, in contrast to the 
 analytical problems raised by the loss of conformal invariance.
 Extensions to general types of geometry and the inclusion of 
 a mass term in the theory should also be within reach of the method. 
 Optical systems in $2D$ and $3D$ with alternative Lagrangians,
 and realistic phase transitions can also be subject to the same techniques.
 Finally, we should stress that since in general we are able to 
 compute the full set of Bogolugov coefficients, we do not have to
 restrict the initial state to be in the vacuum. Time evolution of
 multi-particle states such as thermal configurations, can be easily
 simulated within this framework.

\acknowledgments

 N. D. A. was supported by PPARC.
 N. D. A. would like to thank Claudia Eberlein and Fernando Lombardo 
 for useful suggestions. Part of the research was done within the
 framework of the E.S.F. COSLAB network.
     
\begin{appendix}
     
\section{Numerical Algorithm for Moving Boundary Conditions}
\label{appA}

 In all simulations discussed above the scalar field wave equation
 was solved using a straight-forward Leap Frog
 algorithm (see for example \cite{Numerical_Recipes}). Only the
 case of the moving mirror (Section ~\ref{secIII}) requires more
 attention. For a box with time-dependent length, if the lattice
 spacing is kept constant, the number of points
 in the simulation domain must change and this may lead to large
 numerical errors. 
     
 The Leap Frog (LF) discretization of the wave equation
 Eq.~(\ref{EOM_moving_mirror}) is
     
 \begin{eqnarray}
 & & \Pi_{i,n+1/2}=\Pi_{i,n-1/2} + dt\, \nabla^2 \phi\mid_{i,n}\nonumber \\
 & & \phi_{i,n+1}=\phi_{i,n} + dt\, \Pi_{i,n+1/2},
 \label{LF}
 \end{eqnarray}        
 where $i$ is the spatial lattice index and $n$ the time step. 
 As usual with LF-type algorithms, the field and its momentum are
 defined respectively on integer and half-integer time-steps.
 As before the coordinate of the moving boundary will be denoted
 by $X(t)$, and at each time-step its value will not necessarily
 lie exactly on one of the lattice spatial points. This discrepancy
 will have to be taken into account. In particular, as $X(t)$
 increases new points will be introduced in the lattice and one must
 find a rule for assigning a field or momentum value to these.
   
\begin{figure}
 \centerline{\psfig{file=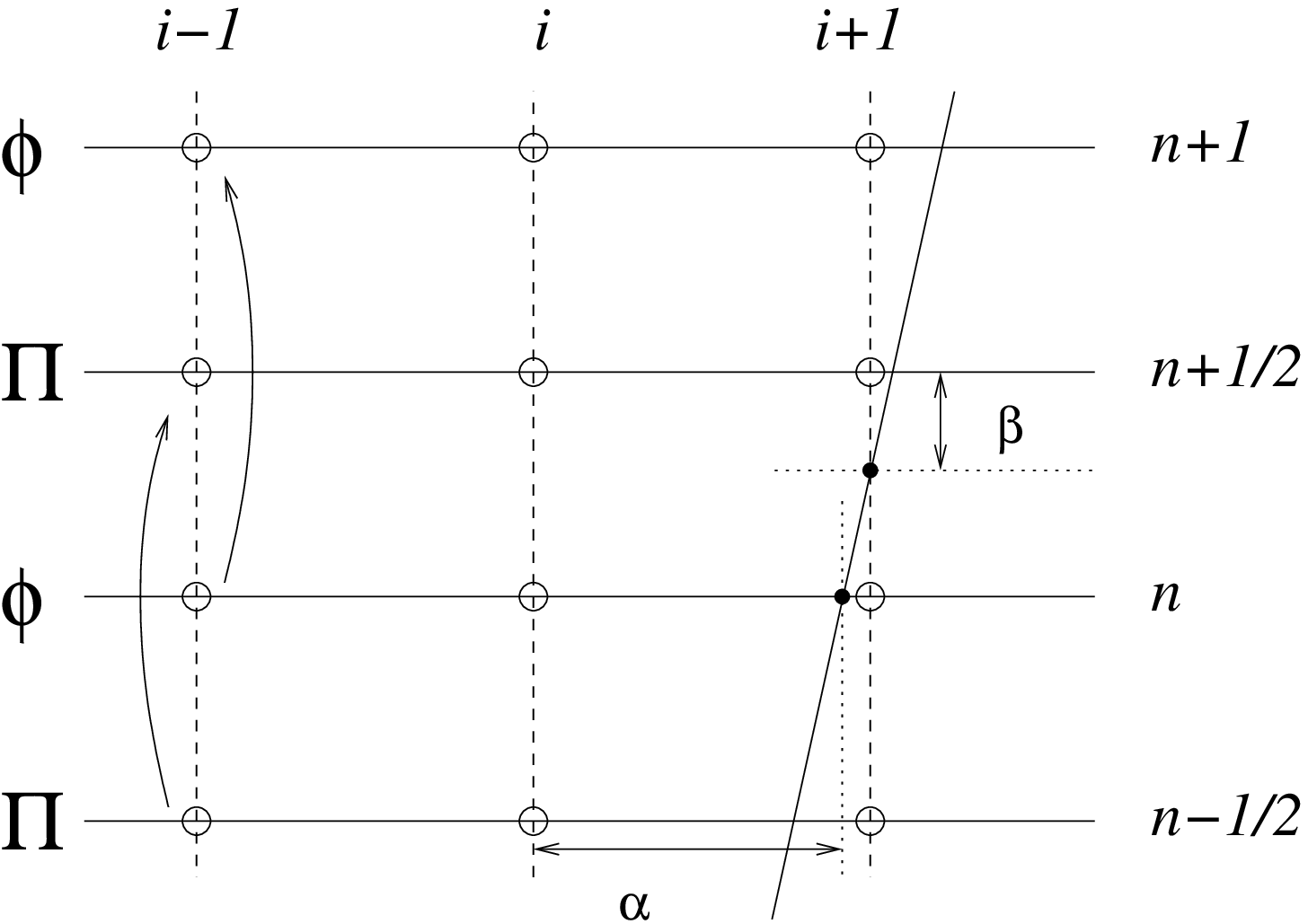,width=3.25in}} 
 \caption{Update algorithm for the moving boundary problem:
 integer $i$ denotes spatial lattice index and  $n$ is the time-step.
 For higher accuracy, we define auxiliary sites in both space and time 
 directions as the boundary's trajectory - diagonal line - crosses 
 the lattice.
 The separation between these and their adjacent integer lattice 
 coordinates is denoted respectively, by $\alpha$ and $\beta$.} 
 \label{method}
 \end{figure}

 Let us start with the case where the new lattice point $i+1$ lies in
 a half-integer time slice $n+1/2$, as depicted in Fig.~\ref{method}.

 We start by differentiating in respect to time the boundary condition
 $\phi(X(t),t)=0$,
 \begin{equation}
 \frac{\partial \phi}{\partial x}\mid_{(X(t),t)}\dot{X}(t)+\Pi(X(t),t)=0
 \end{equation} 
 This allows us to calculate the momentum at the boundary in terms
 of the spatial derivative of the field. Since the field vanishes at
 the boundary, we can approximate $\partial_x\phi$
 at the point $(i+1,n+1/2)$ by $-(\phi_{i,n}+\phi_{i,n+1})/(2\,dx)$,
 using the fact that $\phi_{i,n+1}$
 can be evaluated in advance. This leads to:
 \begin{equation}   
 \Pi_{i+1,n+1/2}=
 \frac{\dot{X}_{n+1/2}}{2\,dx}(\phi_{i,n+1}+\phi_{i,n}) 
 \label{beta_0}
 \end{equation} 
 Note however that here we implicitly assume that the boundary 
 crosses the lattice at precisely the point $(i+1,n+1/2)$, which is
 not necessarily true. The above estimate can be improved 
 by introducing an auxiliary fictitious lattice point $(i+1,n+1/2-\beta)$, 
 the exact location where the mirror's trajectory intersects the
 lattice. The time difference between the crossing and 
 time-step $n+1/2$ is given by $\beta\, dt$ and can be evaluated by
 interpolating the boundary trajectory between steps $n$ and $n+1/2$. 
 At the crossing point we have
 \begin{equation}
  \Pi_{i+1,n+1/2-\beta}=\frac{\dot{X}_{n+1/2}} {2\,dx}
  \left[ (1-2\beta)\, \phi_{i,n+1}+(1+2\beta)\,  \phi_{i,n} \right]
 \label{beta}
 \end{equation}
 Note that by setting $\beta=0$ we recover the previous estimate 
 Eq.~(\ref{beta_0}).
 Using Eq.~(\ref{beta}) we can then update the field at time-step $n+1$
 as
 \begin{equation}
  \phi_{i+1,n+1}=dt\,(1/2+\beta)\,\Pi_{i+1,n+1/2-\beta}
 \label{nphi}
 \end{equation}
 The next momentum up-date, at $n+3/2$, will be calculated 
 taking into account that boundary was placed at $(i+1,n+1/2-\beta)$.
 
 This method is easily generalised to the situation where the newly
 introduced lattice point lies on a integer time slice $n+1$. 
 We proceed in a similar way, defining an auxiliary momentum point 
 at $(i+1,n+1/2-\beta)$, though in this case $\beta$ will be negative.
 The following updates proceed exactly in the same way as described
 in Eqs.~(\ref{beta}) and (\ref{nphi}).
     
  When the trajectory of the mirror is such that no new points
 need to be introduced, we must still take some care when
 updating the fields near the boundary. In particular, the calculation
 of the second derivative term $\nabla^2 \phi\mid_{i,n}$ in Eq.~(\ref{LF})
 should reflect the fact that the boundary may lie between lattice
 sites. Referring again to Fig.~\ref{method} we define $\alpha$ as the
 distance in lattice units from point $(i,n)$ to the mirror. 
 Since the field vanishes at the boundary we will have 
 \begin{equation}
  \nabla^2 \phi\mid_{i,n}=\frac{\alpha \phi_{i-1,n}-
                 (1+\alpha)\phi_{i,n} }{\alpha\,dx^2}
  \label{alpha}
 \end{equation}
 By setting $\alpha=1$, this expression reduces to the less accurate
 estimate for the second derivative, where the boundary's position
 is identified with lattice point $i+1$.
    
 \begin{figure}
 \centerline{\psfig{file=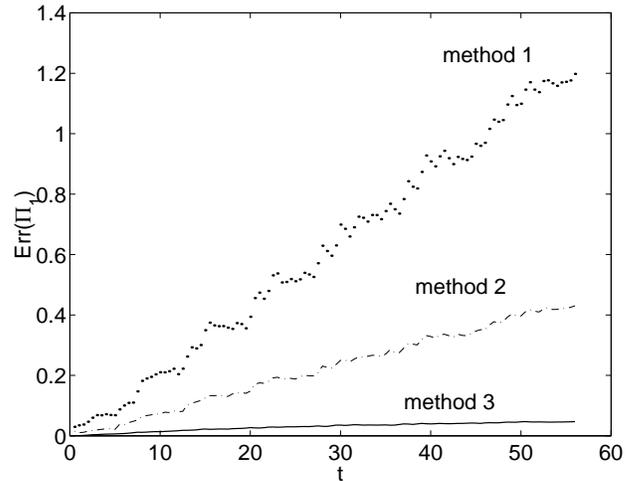,width=3.25in}} 
 \caption{Numerical errors for three different methods for the
  first component of the momentum field $\Pi_1$. 
  A single mode $\omega=n \pi/X_0$ with $n=16$ was evolved and compared
  with the corresponding analytical solution. The cavity expanded
  with constant velocity $v=0.9$ from initial size $X_0=50$.} 
 \label{errors}
 \end{figure} 
    
 In order to check the accuracy of the algorithm, we have used the 
 exact analytical solution Eq.~(\ref{basis uniform}) for the uniformly
 expanding box and compared it to the corresponding numerical
 solution. In Fig.\ref{errors} we show for a particular solution,
 the numerical errors for three different methods based on the
 approximations discussed above. The error is defined as the 
 (normalised) integral of the square of the difference between the
 exact and numerical solutions.
 Algorithm $1$ is obtained by setting $\beta=0$ and $\alpha=1$ 
 in Eqs.~(\ref{beta}) and (\ref{alpha}). This corresponds to a straight
 forward implementation of the Leap Frog method with no
 smoothing at the boundary in either time or space directions. Method
 $2$ includes the corrections for the momentum at the boundary 
 but non-improved spatial derivatives ($\alpha=1$). Finally, in method
 $3$ both $\alpha$ and $\beta$ are evaluated at each time-step
 according to the mirrors's position. The improvement in accuracy is clear:
 algorithm $3$ decreases the errors by a factor of $10$ over method $2$
 and by a factor of $25$ over method $1$. For the purposes of the
 simulations in this work this is enough. Direct observation of the
 numerical solutions shows that the moving boundary introduces high-frequency
 discontinuities in the fields (in particular in the momentum). These
 are considerably reduced by method 3, and in the end they tend to
 average out when the Bogolugov coefficients are evaluated, leading
 to reasonably small errors in the final physical quantities.
          
\end{appendix}

\end{document}